# Effect of scattering and electronic noise upon selection of detectors for Gamma Computerized Tomography

Kajal Kumari[1$], Snehlata Shakya[1], Mayank Goswami[1#],

[1]Divyadrishti Laboratory, Department of Physics, IIT Roorkee, Roorkee, India
Department of Physics, IIT Roorkee, Roorkee, India

[$]kkumari@ph.iitr.ac.in and [$]kajalkumari7867@gmail.com

[#]mayank.goswami@ph.iitr.ac.in,

## Abstract

Computed tomography (CT) has become a vital tool in a variety of fields as a result of technological developments and continual improvement. High-quality CT images are desirable for image interpretation and obtaining information from CT images. A variety of things influence the CT image quality. Various research groups have investigated and attempted to improve image quality by examining noise/error associated with CT geometry. This study aims to select detectors for CT, which yield the least amount of noise in projection data. Three distinct gamma-ray detectors that are routinely used in CT have been compared in terms of scattering and electrical noise. The sensitivity of Kanpur Theorem-1 to scattering noise is demonstrated in this work and used to quantify the relative level of scattering noise. The detector measures the signal multiple times, and the standard deviation of the signal is used to calculate the electronic noise. It is observed that IC CsI(Tl) scintillation detector produces low electronic noise and relative scattering noise as compared to conventional electronic detectors; NaI(Tl) and HPGe.

Keywords: Gamma radiation, Scintillation detector, Semiconductor detector, Computerized tomography, electronic noise, Radiation scattering.

## Introduction

### Electronic noise

Computed tomography (CT) is an established nondestructive evaluation tool. The CT image quality is characterized in terms of image contrast, spatial resolution, image noise, and artifacts in the recovering images[1]. Electronic noise and scattering radiation predominately result in image noise and artifacts[2], [3]. Electronic noise is primarily caused by: (a) noise from the integrating amplifier due to higher feedback capacitance; and (b) quantization noise caused by an analog to digital converter (ADC) over a broad signal range[4]. Noise becomes comparable to the signal if a relatively weak radiation source w.r.t (a) distance between source and detector and/or (b) object (made of high Z value) is used.

The former point is important in the field of medical diagnosis. Additional dose reduction reduces the probability of passive cancer and vascular disease development[5]–[7]. Low radiation dose measurement data may have electronic noise comparable to signal amplitude infecting/corrupting the CT examination's diagnostic value. The noise reduction technique is an ongoing area of research limitations. The noise reduction techniques use: (a) statistically rich data measurement methods, (b) filter-based approach, (c) modification of underlying models, and (d) advanced hardware.

Measurement data with the best Gaussian fit with least standard deviation (SD) and low value of full-width half maxima (FWHM) is preferred and shown to give the best reconstruction quality[8]. Most of the post-processing noise filtering approaches may be applied during three stages: (a) before reconstruction [4], [9] (b) while solving the inverse problem [10], and/or (c) after reconstruction on pre-final image[11], [12]. Several modifications are proposed in beer-lambert law, the basic attenuation model for variance-related noise reduction [13]. As far as using advanced hardware is concern, detector and associated electronics are major components.

Several comparison studies are provided preferring Integrated Circuit (IC) detectors over detectors with distributed electronics as far as reduction of electronic noise is concern[3], [14], [15]. It has been observed that image noise can be reduced by approximately 10% using the IC detector[15]. The CT experiment is performed, first time, using distributed electronics detectors. The same CT experiment is performed using IC detectors but this time lowering the radiation dose. It has been shown that a dose reduction of 20% can be accomplished to match standard deviation (SD) of CT number / pixel value obtained using distributed electronics detector[15]. The impact of electronic noise upon reconstructed image quality is quantified by the standard deviation of CT numbers (SDCT) in the uniform region of interest (ROI), and artifact level is quantified by the square root of the difference in the square of SDCT between streak artifacts regions and adjacent artifact-free region. The CT experiment is performed to evaluate image noise in homogeneous phantom (cylindrical water phantom) and patients, as well as artifact in semi-anthropomorphic phantom[3]. Use of SDCT parameter as image noise estimation can be justified if true images is available which is not possible in case of clinical studies.

## Nonlinear Scattering of radiation

Another critical source of error (w.r.t. transmission CT) is inability to remove the component related to photon scattering inside: (a) nuclei, (b) the crystal of the detector, and (c) the object, from the projection data. The latter component can further be divided into two components: (i) scattering taking place inside the radiation beam originated from source and extended to the end of the detector head, and (ii) outside of this beam. The former component can be included in modelling by estimating the weight factors in pixel by beam. The scattering of radiation affects the measured data quantitatively and qualitatively[16], [17]. To the best of our knowledge the former two phenomenon (a) and (b) are not investigated. Scatter correction models aren't commonly employed in Gamma computed tomography, but they're popular in single photon emission computed tomography (SPECT), positron emission tomography (PET), and X-ray computed tomography. Several models and methodologies are developed in SPECT, PET and X-ray tomography to suppress or estimate and then eliminate the photon scattering from the measured data[18]–[26] for latter phenomenon.

The detectors employed in this research are of different diameters and crystal thicknesses. The increase in crystal size raises the photo fraction (ratio of counts under complete energy peak to total counts in the energy spectrum), owing to a higher contribution from Compton scattering as well[27].

This study utilizes a gamma CT system to evaluate the noise level in counts measured by three different gamma detectors and its impact on CT images. Exact quantification of error in reconstruction is only possible if prior information of the scanned object is available. If such information is already available would render the non-destructive evaluation unnecessary. Kanpur theorem, however, estimates the inherent error level in projection data without having prior information of the scanned object[28]. In our work, performance evolution tools based on with or without the prior knowledge of scanned object are used, which are Sorensen-dice similarity coefficient[29] and Kanpur theorem[30]. Until now, Kanpur theorem -1 (KT-1) has been employed to calculate the inherent noise in projection data of CT. This is the first time that the sensitivity of KT-1 to error caused by scattering radiation has been demonstrated. In this study, KT-1 is used to estimate the *relative* scattering noise level in projection data obtained by three gamma detectors. To estimate the electronic noise level, the standard deviation parameter is used.

# Motivation

To build an advanced CT hardware, one needs to trade-off between overall cost, robustness character, power consumption, size, and noise level to select a detector and its associated electronics. This work presents a methodology for the same.

# Materials & Methods

### 1. Electronics Noise estimation methodology

A detector is placed in front of the source at 42cm (same as in case of CT) but without any object in between. Radiation counts are measured multiple times. Details of electronics setting are same as set in CT measurement except window setting. Full spectrum energy range is set in this case. It is estimated that the gamma source emission has insignificant energy dependent randomness (maximum SD 2) as compared to randomness due to electronic noise. This calculation is explained in [supplementary file](). Scattering of photon inside the nuclei will be same for all detectors. Different types of detectors will have different crystal materials, size, and thickness. Scattering inside a given crystal or radiation source nuclei is dependent on material distribution and will have insignificant randomness in multiple measurement data. It will not contribute to component that imparts randomness so the value of standard deviation in total counts will give an estimate for the electronic noise only.

### 2. Methodology to estimate relative error due to scattering

CT projection data is gathered using single detector by following an innovative CT scheme. Three different energy ranges: (a) gamma photo peak region, (b) Compton edge and Compton valley, and/or (c) full energy spectrum are used. This strategy has generated 3 projection data for each detector, in total 9 data sets.

It is expected that CT projection data from latter energy range (full energy spectrum range settings) will contain maximum error component related to scattering. Similarly, CT projection data from gamma photo peak is expected to contain least scattering error.

Kanpur theorem is used to estimate the inherent error in these projection data sets. Detail for this error estimate is given in [supplementary file](). *This inherent error includes electronics noise, and modelling error due to scattering component in source nuclei, and detector crystal.* Since single detector is used it will not be able to account for the scattering component outside of the radiation beam (explained earlier). Existing transmission tomography methods can only include the scattering component which is inside the radiation beam; however, all components are present in the CT measured data.

The details and properties of different types of gamma detectors are given in next subsection. The details of electronics parameters settings and CT geometry parameters are given in subsection 3 and 4, respectively. The details of the phantom used in gamma CT is provided in fourth subsection. The 'Measurement' subsection discusses the measurement scheme to evaluate the *relative* scattering noise and electronic noise. CT image reconstruction is provided in last subsection 'Image reconstruction and noise evaluation'.

### 3. Detectors and electronic settings

To evaluate the level of electronic and scattering noise in a given detector following factors may be considered: crystal thickness, crystal material type, and kind of electronic circuit (distributed and integrated). We have selected three distinct gamma-ray detectors: an compact CsI(Tl) Multi Pixel Photon Counter scintillation IC detector, two detectors with distributed electronics: (a) a high-resolution HPGe semiconductor detector, and (b) a NaI(Tl) scintillation detector [31], [32]. The specifications of these three detectors are mentioned in table 1. The latter set of hardware (make: electronics enterprise Ltd. India) has NaI (Tl) scintillator crystal is coupled with its photomultiplier (PM) tube; Its anode output is amplified by the separate amplifier circuit. It is finally connected to a PC via a single-channel analyzer (SCA). The HPGe detector (make: Canberra and NI-PIXI-1031) is interfaced to Nuclear Instrumentation Module (NIM) and a multichannel pixie-4 data acquisition system. The third CT setup (make: Hamamatsu model C12137) contains CsI (Tl) scintillation crystal coupled with a photodiode, preamplifiers, and analog-to-digital converters all integrated into the same silicon chip. Each of these have their own manufacturer provided data acquisition software.

The PM tube of the NaI(Tl) detector is subjected to a high voltage of 650V[8], whereas of HPGe detector is subjected to a high voltage of 4500V. The CsI (Tl) detector requires USB to get connected with a PC in plug-in-play mode. The counting time is set at 30 seconds[8] for each projection data set. Due to low activity of the radioactive source, and for selected geometry, we have observed that 30 seconds is insufficient for counting gamma radiation by CsI (Tl) detector in our experiment. Therefore, in the CsI detector program, the next option for counting time is 1 minute, followed by 10 minutes; we chose 10 minutes. The counts are normalized for all three detectors at the end.

## 4. CT Geometry and Phantom

The gamma CT experiment comprises an encapsulated radioactive source Cs-137 of activity 1.5µCi, phantom, and gamma-ray detectors. For scanning purposes, a well-known cylindrical-shaped phantom is used. The phantom is made up of Perspex material with a diameter of 12 cm. Two holes, 3.8 cm and 0.8 cm in diameter are drilled and afterward filled with concentric aluminium and iron cylinders.

Gamma CT experiment is performed using three types of gamma-ray detectors NaI(Tl), HPGe, and CsI(Tl), shown in figs. 1(a), 1(b) and 1(c). It uses fan-beam geometry shown in Fig 1(d) with a fan beam angle of 37.1°.

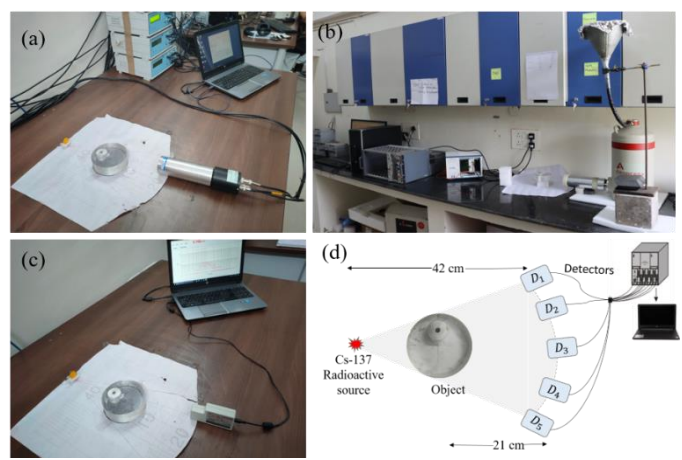

Fig.1: Gamma CT measurement Systems(a) NaI(Tl), (b) HPGe, (c) CsI(Tl), (d) Schematic

Source to detector and object to detector distance is kept at 42cm and 21cm, respectively. The projection data is obtained by rotating the phantom in 18 steps, moving the detector in five steps in an 'arc' form because only one detector of each kind is available. The 'arc' must be formed in such a way that first and last detector receives spread of radiation passing via object. Since we have one detector of each type, and its not advisable to move HPGe detector, scheme2b is used to create four virtual detectors to complete the 'arc' of five and create CT data from 5 detectors for 18 view/projections.

We have compared the sinogram taken by this *innovative* scanning method, scheme2b with the scanning scheme2a and conventional scheme1 (when we would had used five detectors same time). Please note that we do have 5 NaI (Tl) detectors but for uniformity sake, we only used single NaI(Tl) for CT. It is found that in all three scanning schemes give same sinogram. Description of these three scanning schemes is given in [supplementary file](supplementary file).

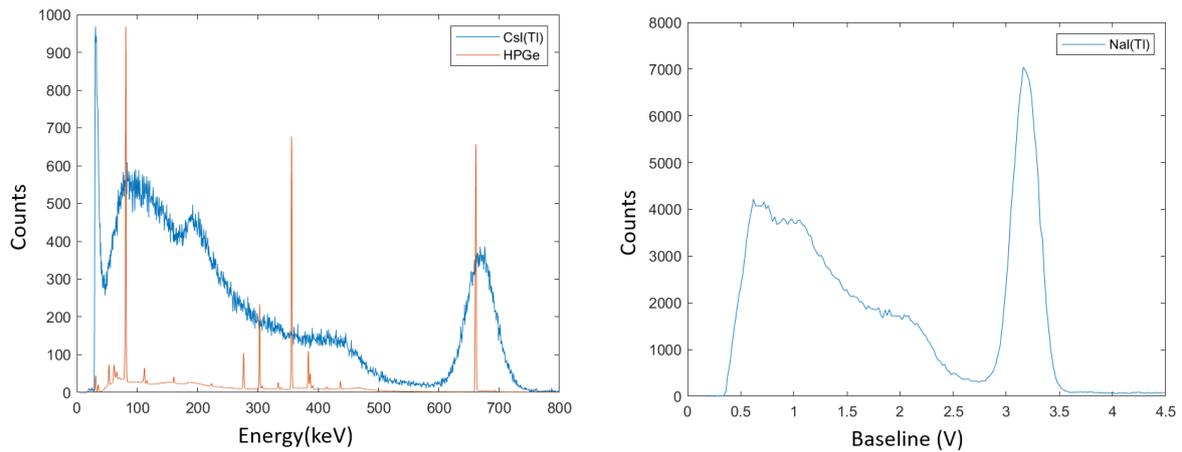

Fig.2: Energy spectrum of Cs-137 radioactive source detected by CsI(Tl), HPGe and NaI(Tl) detector.

## 5. Measurement

Gamma CT experiment is performed using one type of gamma-ray detector setup at a time. The same CT geometry is used for the rest of the detectors.

Three energy ranges are selected to estimate the relative scattering noise in CT projection data obtained by three different gamma detectors: (a) gamma photopeak counts, (b) Compton edge and Compton valley, and (c) full energy spectrum. The full energy spectrum is recorded, using multi-channel analyser (MCA) with HPGe detectors. CsI(Tl) detector has on-board / integrated MCA on its electronics chip. The counts for particular (as desired) energy range are extracted out of this data later. The spectrum from NaI(Tl) is measured using its available SCA. The energy window selected to contribute the counts in projection data from different energy range is given in table 1.

Table 1: Specifications of three gamma detectors

| Properties of detectors | NaI(Tl) | CsI(Tl) | HPGe |
|---|---|---|---|
| Crystal Diameter | 25.4 cm | 13 cm | 55.60 cm |
| Crystal thickness | 50.8 cm | 20 cm | 60 cm |
| Hygroscopic | Yes | Slightly | Cooling required |
| Power Consumption | +650V | +5V | -4500V |
| Cost (in Lacs): single detector + electronics | 0.93 + 54 | 3.2 (inclusive) | 35 + 15 |
| Estimated Resolution | 8.9% | 8.2% | 0.23% |
| Gamma photopeak energy window (selected) | 2.8 to 3.52V | 610 to 762keV | 652 to 669keV |
| Compton energy window (selected) | 1.5 to 2.7V | 357 to 552keV | 350 to 550keV |

Due to the lack of an integration mode in SCA, threshold mode is used to measure the counts throughout the full energy spectrum. The minimal threshold is established at 0.3V because below than this value zero counts obtained. The threshold value for the CsI(Tl) and HPGe detectors is chosen 30 keV and 35 keV, respectively. Theses default settings are used for CsI(Tl) and HPGe detector. Individual detectors' energy spectrums are visualized to select energy ranges. The measured energy spectrum from all three detectors is shown in fig.2.

## 6. Image reconstruction

KT-1 estimates the inherent noise in projection data if reconstructed using the convolution back-projection algorithm (CBP) for parallel beam geometry[33]. We first transformed the fan-beam projection data into a parallel beam equivalent to utilize the Kanpur Theorem. Apt values of input information such as geometry information, number of detectors, number of rotations, and type of filter functions are utilized to develop custom reconstruction codes.

In our work, we didn't use the RMSE of reconstructed image w.r.t. cyber phantom as an error estimation parameter. Analysis in support is given in supplementary file[34]. The Sorensen dice coefficient is employed for the first time in CT to compute the similarity in reconstruction with cyber phantom to estimate error.

## Results

Electronics Noise estimation methodology (described above) is applied using three detectors (details mentioned above).

Figure 3 depicts the SD of NaI(Tl), CsI(Tl) and HPGe detector. Lowest SD is observed for CsI(Tl) detector and highest for NaI(Tl) detector.

$GOF_{RMSE}$ of KT-1 signature is obtained. The bar graph of $GOF_{RMSE}$ is shown in fig. 4. If we compare three energy regions, it is observed that noisy projection data is obtained in full energy spectrum region and least noisy projection data in gamma photopeak region for all three gamma detectors.

We also compared the gamma detectors used in this study. It is observed that regardless of the energy region, least noisy projection data is obtained for CsI(Tl) detector and noisy projection data is obtained for NaI(Tl) detector. For full spectrum and Compton region, NaI(Tl) and HPGe detectors gives comparative $GOF_{RMSE}$ values. While a significant difference in $GOF_{RMSE}$ value can be seen for gamma photopeak region. CsI(Tl) gives least noisy but a comparative value of $GOF_{RMSE}$ for gamma photopeak region is also obtained by HPGe detector.

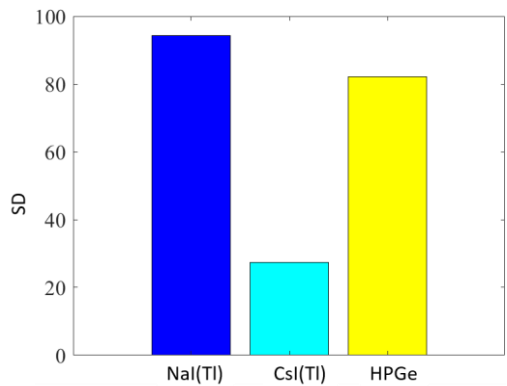

Fig. 3: Standard deviation of NaI(Tl), CsI(Tl) and HPGe detector.

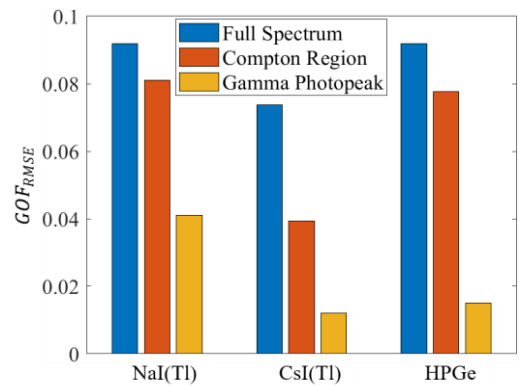

Fig. 4: $GOF_{RMSE}$ of KT-1 signature for $N_{avg}$ value.

The CT images for gamma photopeak region obtained by three gamma detectors is shown in fig. 5. The CsI(Tl) detector has the highest dice similarity coefficient, followed by the NaI(Tl) and HPGe detector. Dice similarity coefficient's result matches with human perception i.e. visualization of reconstructed images w.r.t cyber phantom. In case of NaI(Tl), and HPGe detector, first and last column in reconstructed image contains no valuable information. This factor is also included in human perception.

This version is uploaded on arXiv on 25 May 2022

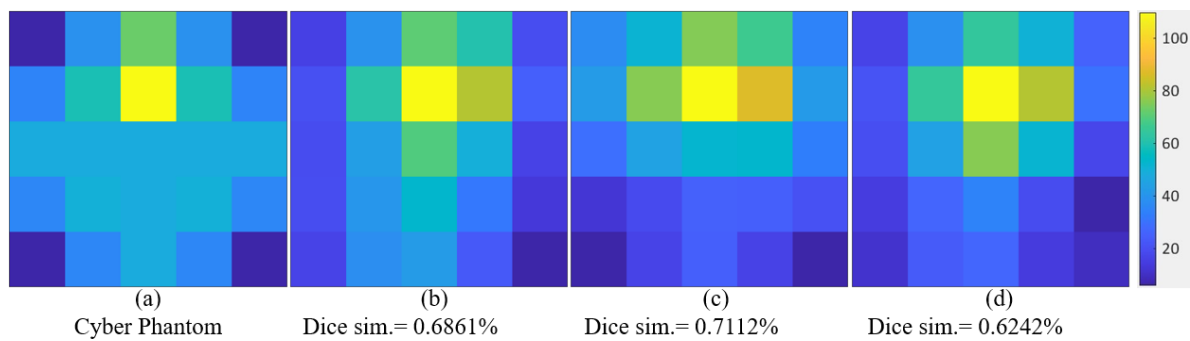

(a) Cyber Phantom   (b) Dice sim.= 0.6861%   (c) Dice sim.= 0.7112%   (d) Dice sim.= 0.6242%

Fig.6: (a) Cyber phantom, CT reconstruction result of gamma photo peak region detected by (b) NaI(Tl), (c) CsI(Tl), and (d) HPGe detector.

Figure 7 shows the normalized sinograms / projection data measured from all three detectors and obtained from the cyber phantom. The sinogram of cyber phantom shows distinct patches with uniform gradient / variations. The Fig. 7(b) has highest difference in these regions. HPGe has least difference in first half of the second row. CsI(Tl) has best similarity with cyber phantom for first half of the last row and last patch of the first row However, more or less all four have similar visual characteristics.

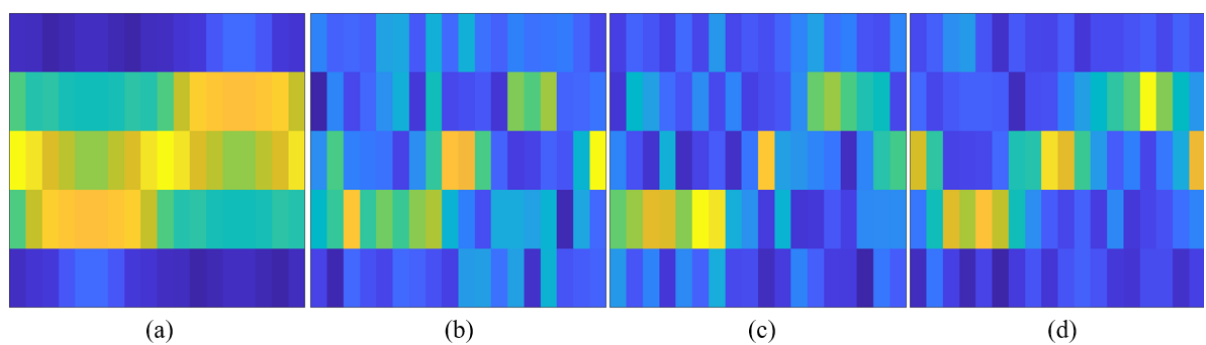

(a)   (b)   (c)   (d)

Fig. 7: Sinogram of (a) cyber phantom, (b) NaI(Tl), (c) CsI(Tl), and (d) HPGe

## Discussion and Conclusion

The following are the major findings of this study:

1) IC detector has low level of electronic noise as compare to detectors with distributed electronics.
2) For the first time the sensitivity of KT-1 to error due to scattered radiation is demonstrated and it is utilized to evaluate the relative scattering noise in three gamma detectors.
3) It is observed that out of three gamma detectors, CsI(Tl) detector has least inherent noise in projection data. The integrated electronic circuit, as well as the crystal size might be reason for least value of electronic noise and scattering radiation.
4) The influence of electronic noise and scattering noise on CT images is illustrated.
5) To minimize the scattering radiation w.r.t transmission tomography one needs to consider crystal diameter and thickness.
6) It is concluded that the IC CsI scintillation detector produces the lowest electronic and scattering noisy projection data. Based on this research, it is advised to prefer compact IC CsI(Tl) detector over NaI(Tl) and HPGe (expensive and requires additional maintenance) detector for better CT image quality.

## Acknowledge

K. Kumari is thankful to the Council of Scientific and Industrial Research (CSIR), India for the fellowship. M. Goswami would like to acknowledge to Prof. Anil K. Gourishetty, Prof. Ajay Y. Deo and members of the RDS Lab at Department of Physics, IIT Roorkee for allowing to use HPGe detector system.

## Credit authorship contribution statement

K. Kumari: Methodology, Experiment, Data Acquisition, coding and Writing. M. Goswami: Methodology, Funding, Supervision, Writing.

## Conflicts of Interest: The authors declare no conflict of interest.

## Supplementary Data:

## Estimation of Randomness of radiation emission by Source:

SD of repeatedly measured counts can be calculated in two ways: (1) The SD of total counts in full energy spectrum, and/or, (2) the energy dependent SD of counts. The former methodology is used to estimate the level of electronic noise. The latter one is used to estimate the randomness of radiation emission by source. SD of counts at a particular energy is calculated. The same is done for all energy values. In this way, the spectrum of energy dependent SD is obtained, which shows that maximum SD is obtained for X-ray peak which is 2 while for gamma photopeak 0.6 SD is obtained.

## KT Theorem:

In KT-1, the plot between the reciprocal of gray level value of reconstructed image and the second derivative of filter function W″ (0) can be used to indicate noise in CT images. The window/filter functions and their second derivative value implemented in this work is given in table 2. The linearity behaviour of plot shows that the projection data is noise-free or has a low noise level. The root mean square error of the linear fitting $GOF_{RMSE}$ is used as a parameter to check the linear behaviour of KT-1 signature.

Table 2:

| Code | Class | Nature | W″ (0) |
|------|---------|--------|--------|
| h99 | Hamming | Sharp | 0.001 |
| h91 | Hamming | Sharp | 0.083 |
| h75 | Hamming | Medium | 0.250 |
| h54 | Hamming | Smooth | 0.460 |
| h50 | Hamming | Smooth | 0.500 |

## Scanning Schemes description is given below:

### Scheme1:

When 5 detectors are available then the experiment can be performed in following steps:

1. Draw the CT geometry (the position of source, position of object w.r.t source, angular distance and angular position of five detectors in an 'arc' w.r.t source) on table/platform/paper.
2. Place the source at its position.
3. Place the object at its position, by keeping its center coincide with the center of source.
4. Place the five detectors at their respective angular position on 'arc'.
5. Rotate the object (without changing its center) in 18 steps until a complete rotation while recording their corresponding counts in all five detectors.

### Scheme2a:

When detector can move then CT Scanning is performed in following steps:

1. Draw the CT geometry (the position of source, position of object w.r.t source, angular distance and angular position of five detectors in an 'arc' w.r.t source) on table/paper.
2. Place the source at its position.
3. Place the object at its position, by keeping it's center coincide with the center of source.
4. Place the single detector at first angular position on 'arc'.
5. Rotate the object (without changing its center) in 18 steps until a complete rotation while recording their corresponding counts.
6. Change the angular position (on arc) of detector w.r.t source and repeat the (5) point.
7. Repeat from point (6) until we get data from five angular position of detectors placed in an 'arc'.

## Why RMSE is not accurate error estimate:

Fig. 1 shows that minimum RMSE is obtained for NaI detector while maximum RMSE is obtained for CsI (Tl) detector. However, human visualization of these CT images with cyber phantom conclude that the CT image obtained by CsI (Tl) match closely compare to HPGe and NaI(Tl) detector's CT images. It demonstrates that RMSE result doesn't match with human visualization.

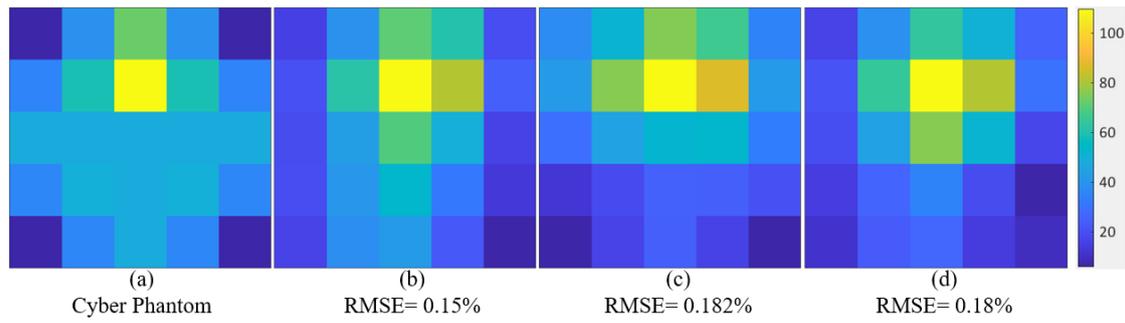

Fig. 1: (a) Cyber phantom, CT reconstruction result of gamma photo peak region detected by (b) NaI(Tl), (c) CsI(Tl), (d) HPGe detector.